\documentclass{emulateapj}
\def\stacksymbols #1#2#3#4{\def\theguybelow{#2}
        \def\verticalposition{\lower#3pt}
        \def\spacingwithinsymbol{\baselineskip0pt\lineskip#4pt}
        \mathrel{\mathpalette\intermediary#1}}
\def\intermediary #1#2{\verticalposition\vbox{\spacingwithinsymbol
        \everycr={}\tabskip0pt
        \halign{$\mathsurround0pt#1\hfil##\hfil$\crcr#2\crcr
                \theguybelow\crcr}}}
\def\lta{\stacksymbols{<}{\sim}{2.5}{.2}}
\def\gta{\stacksymbols{>}{\sim}{3}{.5}}

\begin{document}

\title{BARYONICALLY CLOSED GALAXY GROUPS}

\author{ William G. Mathews\altaffilmark{1}, 
Andreas Faltenbacher\altaffilmark{1}, 
Fabrizio Brighenti\altaffilmark{1,2} \&
David A. Buote\altaffilmark{3}}

\altaffiltext{1}{University of California Observatories/Lick Observatory,
Department of Astronomy and Astrophysics,
University of California, Santa Cruz, CA 95064~mathews@ucolick.org~
fal@ucolick.org}

\altaffiltext{2}{Dipartimento di Astronomia,
Universit\`a di Bologna,
via Ranzani 1,
Bologna 40127, Italy~fabrizio@aspera.bo.astro.it}

\altaffiltext{3}{Department of Physics and Astronomy, 
University of California, Irvine, CA 92697~buote@uci.edu}

\begin{abstract}
Elliptical galaxies and their groups having the largest 
$L_x/L_B$ lie close to the locus 
$L_x = 4.3 \times 10^{43} (L_B/10^{11}~L_{B\odot})^{1.75}$
expected for closed systems having baryon fractions 
equal to the cosmic mean value, 
$f_b \approx 0.16$. 
The estimated baryon fractions for several 
of these galaxies/groups are also close to $f_b = 0.16$
when the gas density is extrapolated to the virial radius.
Evidently they are the least massive baryonically closed systems.
% and their hot gas 
%X-ray luminosities $L_x$ 
%should not be significantly surpassed by future observations. 
Gas retention in these groups implies that 
non-gravitational heating cannot exceed about 1 keV per particle, 
consistent with the heating required to 
produce the deviation of groups from the 
$L_x - T$ correlation for more massive clusters.
Isolated galaxies/groups with X-ray luminosities significantly 
lower than baryonically closed groups may have undermassive 
dark halos, overactive central AGNs,
or higher star formation efficiencies.
The virial mass and hot gas temperatures 
of nearly or completely closed 
groups correlate with the group X-ray luminosities and the optical 
luminosities of the group-centered elliptical galaxy, 
i.e. $M_{vir} \propto L_B^{1.33}$, 
an expected consequence of their merging history. 
The ratio of halo mass to the mass of the central galaxy 
for X-ray luminous galaxy/groups is $M_{vir}/M_* \sim 80$.
\end{abstract}

\keywords{
galaxies: elliptical and lenticular, CD --
X-rays: galaxies --
galaxies: clusters: general --
X-rays: galaxies: clusters -- 
galaxies: cooling flows
}

\section{Introduction}

Massive elliptical galaxies with similar optical luminosities 
have hot gas 
X-ray luminosities that range over two orders 
of magnitude.
The origin of this scatter, shown in Figure 1, 
has received much attention 
but a full understanding remains elusive. 
There is evidence that gas loss by ram pressure (and tidal) 
stripping has reduced $L_x/L_B$ 
in elliptical galaxies or groups orbiting within 
rich clusters of galaxies 
(Biller et al. 2004; Machacek et al. 2005; Sun et al. 2005).
However, an enormous range in $L_x/L_B$ 
also prevails among non-interacting ellipticals that are isolated or 
at the centers of isolated galaxy groups. 
The correlation between the spatial extent of the X-ray emission 
and $L_x/L_B$ suggests that the driver for this scatter is 
a variation in the virial mass $M_{vir}$ of the halo that surrounds 
otherwise similar elliptical galaxies 
(Mathews \& Brighenti 1998). 
The virial mass $M_{vir}$ and radius $r_{vir}$ are 
found by fitting dark NFW halos 
to the total mass distribution 
derived from X-ray observations of the
hot gas density and
temperature in $10~{\rm kpc} \lta r < r_{vir}$, 
assuming hydrostatic equilibrium.

To gain further insight into the broad range of X-ray emission 
from optically similar galaxies, we draw attention here to 
those ellipticals with the largest X-ray luminosities. 
These isolated galaxy/groups have been variously referred to as 
``[X-ray] overluminous elliptical galaxies (OLEGS)'' 
(Vikhlinin et al. 1999) or ``fossil groups'' 
(Ponman et al. 1994). 
The concept of fossilized groups is meant to imply that 
they are relics of merging among galaxies in a group 
environment, 
although all elliptical galaxies may qualify for this designation.
Jones et al. (2003) provide an operational definition for 
fossil groups in terms of the magnitude difference 
between the first and second brightest group galaxies. 
For our purposes here we simply consider those 
elliptical galaxies with the largest $L_x/L_B$ 
in the $L_x - L_B$ plot, many of which 
have been previously regarded as fossils or OLEGS.
We then note that several of the 
best studied of these galaxies have nearly 
the same baryon mass fraction as the most massive galaxy 
clusters and the {\it WMAP} value, 
$f_b = 0.16$ (Spergel et al.  2003), 
i.e. they appear to be baryonically closed. 
Most baryons are in the hot intragroup gas.
%The optical luminosity 
%$L_{B,E}$ of the central E galaxy 
%correlates with the halo virial mass and with 
%the emission-averaged gas temperature. 

The data in Figure 1 are mostly taken 
from O'Sullivan et al. (2001) (open squares), 
but we have added additional 
X-ray luminous ellipticals assembled from more recent 
observations (filled symbols)
with properties listed in Table 1.
%According to Vikhlinin et al. (1999), 
%the space density of these groups is large, 
%comprising about 20\% of all groups of similar 
%$L_x$ and nearly all field galaxies with 
%$\log L_B \gta 10.80$.
These X-ray luminous 
systems define the upper envelope of the 
luminosity distribution in the $L_x - L_B$ plane. 
While all  estimates of the baryon mass fraction 
$f_b$ require uncertain extrapolations 
beyond the observations to the virial radius $r_{vir}$, 
$f_b$ for several 
X-ray luminous groups in Table 1 indicate
near or complete baryon closure. 
All data have been scaled to $H_0 = 70$ km s$^{-1}$ 
Mpc$^{-1}$. 

\section{Non-Gravitational Heating in 
the NGC 5044 Group}

Galaxy groups and poor clusters with
$M_{vir} \lta 3 \times 10^{14}$ $M_{\odot}$ and $kT \lta 4$ keV
are known to deviate systematically below the 
$L_x - T$ relation
established by more massive clusters, suggesting additional 
non-gravitational energy by cosmic
preheating or AGN activity (as reviewed by Voit 2005).
Consequently, it is remarkable that groups in
Table 1 with $kT \sim 1-2$ keV have survived with
most or all of their baryonic component intact. 

NGC 5044 is a good example of such a group. 
For the purpose of this discussion, 
we have made a preliminary mass model of NGC 5044 
based on gas density and temperature profiles 
observed to $\sim 0.3r_{vir} \approx 300$ kpc from 
Buote et al. (2003, 2004, \& 2006 in prep.).
In the central regions the azimuthally averaged gas density 
$n_{e,obs}$ was replaced with $n_{e,f} = n_{e,obs}/f^{1/2}$ 
where $f(r) = 1 - 0.642\exp(-r_{kpc}/20)$ 
(Buote et al. 2003) is the filling 
factor of the denser gas component at each radius 
responsible for most of the observed emission. 
The model was constructed by first setting the stellar 
parameters -- a de Vaucouleurs profile with luminosity 
$L_B = 4.5 \times 10^{10}$ $L_{B\odot}$, 
effective radiius $R_e = 10$ kpc and stellar mass to light 
ratio $\Upsilon_B = 7.5$ -- that establish the total 
stellar mass $M_{*E} = 3.4 \times 10^{11}$ $M_{\odot}$ 
and potential.
The dark halo is assumed to have an NFW mass profile with 
an adjustable virial mass $M_{vir}$ and concentration 
$c = 433 M_{vir}^{-0.125}$ expected for this mass 
(Bullock et al. 2001).
The equation of hydrostatic equilibrium is integrated for $n_e(r)$, 
fixing the gas temperature $T(r)$ to fit observations 
and extrapolating to larger radii in a $\log T - \log r$ plot.
$M_{vir}$ and the innermost gas density are varied  
until an excellent fit is achieved to the $n_{e,f}(r)$ 
profile throughout the observed region.
The resulting virial mass, $M_{vir} = 4.0 \times 10^{13}$ 
$M_{\odot}$, is similar to our previous estimate 
(Buote et al. 2004) and the virial radius
$r_{vir} = (3 M_{vir}/4 \pi \Delta \rho_c)^{1/3} 
= 841$ kpc with $\Delta = 104$ and $\rho_c = 9.24 \times 10^{-30}$ 
gm cm$^{-3}$.
When the observed gas density profile in NGC 5044
is extrapolated to $r_{vir}$
(Buote et al. 2004; 2006 in prep.),
maintaining the same power law 
$n_e = 0.66r_{kpc}^{-1.45}$ cm$^{-3}$
observed in the region $100 < r < 300$ kpc,
we find that the total gas mass 
is $M_g = 4.8 \times 10^{12}$ $M_{\odot}$,  
in agreement with the mass model.  
The mass fraction in gas is $f_{g} \approx 0.11$.
This corresponds to a baryon ratio 
$f_b \approx f_g/(1 - 0.12) = 0.13$, 
assuming a (conservative) star formation efficiency of $12$\%
(Lin \& Mohr 2004).
At least $80$\% of the initial baryons in NGC 5044 is still bound 
to the group. 
Evidently, the non-gravitational heating received by the gas 
is $\lta 20$\% of the gas binding energy, 
$E_{bind} = 9.6 \times 10^{61}$ ergs. 
\footnote{The binding energy is found by
computing the double integral
$E_{bind} = \int_0^{r_{vir}} \rho_g 4 \pi r^2 dr
\int_r^{\infty} g(r') dr'$
where $\rho_g$ and $g(r) = GM(r)/r^2$ are the gas density
and gravitational acceleration from our mass model
for NGC 5044.}
For simplicity we assume that the percentage difference 
between the observed $f_g$ and the value 
$f_g \approx (1 - 0.12)f_b = 0.14$ expected from WMAP 
is proportional to the amount of non-gravitational 
energy that the gas received as a percentage of $E_{bind}$.

The gas heating efficiency associated 
with accretion onto the 
central black hole in NGC 5044 must be consistent with 
gas retention.  
The central galaxy with mass $M_{*} = 3.4 \times 10^{11}$
$M_{\odot}$ is expected to contain a black hole of mass
$M_{bh} \approx 7.6 \times 10^{-5} M_{*}^{1.12} 
= 6.2 \times 10^8$ $M_{\odot}$ 
(Haering \& Rix 2004). 
During the accretion history of the central black hole,
suppose that a fraction $\eta_h$ of the rest energy $M_{bh} c^2$
heats the intragroup gas, then 
gas retention at the 80\% level suggests 
that $\eta_h \lta  0.2 E_{bind}/M_{bh} c^2 = 0.016$, 
although some of this energy will be radiated away.
The energy radiated by
NGC 5044 during several Gyrs is
$E_{rad} = L_x t = 4.3 \times 10^{60}(t / 5~{\rm Gyrs})$ ergs 
and, since 
no gas is observed to cool below $\sim T_{vir}/3$
in NGC 5044 (Buote et al. 2003),
the minimum accretion heating efficiency is 
$\eta_{rad} \sim E_{rad}/M_{bh}c^2 \sim 
0.004 (t / 5~{\rm Gyrs})$.
Evidently only a tiny fraction of the accretion energy 
released by the central black hole ($\sim 0.1M_{bh}c^2$)
can have heated the intragroup gas in NGC 5044. 
Nevertheless, substantial ongoing AGN-related heating is 
currently observed near the center of the NGC 5044 group
(Buote et al. 2003;
Mathews, Brighenti \& Buote 2004).

The non-gravitational 
energy received by the hot intracluster gas from 
supernovae can be estimated from the 
total mass of iron observed in the NGC 5044 group,
\begin{equation}
M_{Fe} = 0.71 z_{Fe,gas} M_{gas} + 0.71 z_{Fe,*} \sigma M_{*}
\end{equation} 
where the estimated total gas mass is 
$M_{gas} = 4.8 \times 10^{12}$ $M_{\odot}$, 
$0.71$ is the ratio of hydrogen 
mass to total mass 
including helium, and 
the total stellar mass in the group is $\sigma$ 
times larger than $M_{*}$.
The mass-weighted gas iron abundance in the NGC 5044 group 
is $z_{Fe,gas} = 0.16 z_{Fe\odot}$ and we adopt a mean 
stellar abundance $z_{Fe,*} = 0.5 z_{Fe\odot}$ 
and $z_{Fe\odot} = 1.83 \times 10^{-3}$ 
(Grevesse \& Sauval 1998).
The total iron mass can be expressed in terms of 
supernova yields ($y_{II} \approx 0.1$ $M_{\odot}$ and 
$y_{Ia} \approx 0.7$ $M_{\odot}$) and $\eta$, 
the number of supernovae per $M_{\odot}$ of initial stars formed, 
\begin{equation}
M_{Fe} = M_{*i}(\eta_{II} y_{II} + \eta_{Ia} y_{Ia}).
\end{equation}
We assume that none of the supernova iron either 
cooled (Brighenti \& Mathews 2005) or was buoyantly expelled 
to $\sim r_{vir}$. 
If all stars formed at high redshift with a Salpeter 
IMF between 0.08 and 100 $M_{\odot}$ the 
number of stars with 
mass above $8 M_{\odot}$ that become Type II supernovae 
is $\eta_{II} = 0.0068$ per $M_{\odot}$. 
If $\sigma M_{*}$ is the current stellar mass in the group,
the initial mass is $M_{*i} \approx \sigma M_{*}/(1 - \beta)$ 
where $\beta \approx 0.3$ is the fraction of the original 
stellar mass ejected from stars 
(Brighenti \& Mathews 1999). 
If the iron mass $M_{Fe}$ is eliminated between the two 
equations above, we find that the number of Type Ia supernova 
per $M_{\odot}$ of initial stars is 
$\eta_{Ia} \approx 6.7 \times 10^{-4} \sigma^{-1}$.
The total energy released by both types of supernovae is 
\begin{equation}
E_{sn} = M_{*i} (\eta_{II} + \eta_{Ia}) 10^{51}
\approx 1.1 \times 10^{61} ~~~{\rm ergs}
\end{equation}
where we assume $10^{51}$ ergs per supernova and $\sigma = 3$.
Since $E_{sn}/E_{bind} \approx 0.11$ it is possible that 
supernovae energy could eject 
$\sim 10$\% of the baryons from the group. 
But our assumption that all of the SNII energy is communicated
to the hot gas may not be plausible since 
it makes no allowance for radiation losses in SNII remnants.
Of course a flatter IMF 
(e.g. Brighenti \& Mathews 1999;
%Finoguenov, Burkert \& Bohringer 2003; 
%Portinari et al. 2004;
%Tornatore et al. 2004; 
Nagashima et al. 2005) 
could generate enough SNII energy to eject  
20\% of the gas (e.g. Brighenti \& Mathews 2001).  
Clearly, the large baryon fraction $f_b$ observed in 
NGC 5044 and other X-ray luminous groups imposes 
a significant constraint on non-gravitational heating. 

\section{L$_B$ and the Virial Mass}

The X-ray luminosity of the groups with filled symbols 
in Figure 1 along the upper envelope
of the $L_x - L_B$ distribution correlates 
with $L_B$ as 
$L_x \approx 1.7 \times 10^{21} (L_B/L_{B\odot})^{2}$ 
erg s$^{-1}$ (also see Jones et al. 2003).
If these groups are essentially 
baryonically closed, as we propose here, 
then $L_B$ for the group-centered E galaxy 
should also increase with the virial mass.  
This correlation is shown in Figure 2a 
where we plot those 
groups from Table 1 having known estimated $M_{vir}$.
Similar $L_B - M_{vir}$ correlations 
have been found from the 2MASS survey 
(Lin \& Mohr 2004) and 
the galaxy-galaxy lensing data of
Cooray \& Milosavljevic (2005) who find 
$M_{vir} \propto L_B^{1.33}$, which 
agrees well with the group data in Figure 2a 
\footnote{$M_{vir}$ for NGC 6482 
(the leftmost point in Figure 2a) 
may be significantly underestimated since the X-ray 
observations extend only to $\sim 30$ kpc, 
close to the expected transition between the 
influence of the dark halo mass and the much smaller 
stellar mass.}.
The $L_B - M_{vir}$ correlation arises because group-centered 
elliptical galaxies grow by mergers as 
massive satellite group galaxies undergo 
dynamical friction in the dark halos.
Progressively more massive halos contain more mergeable galaxies 
so the final $L_B$ of the group-centered elliptical increases 
with $M_{vir}$ until 
$M_{vir} \gta 10^{14}$ $M_{\odot}$ where dynamical friction 
is reduced by the small galaxy/halo mass ratio.
%(Faltenbacher \& Mathews 2005). 
All baryonically closed groups have been dynamically 
processed in this way. 

The ratio of the total halo mass to that of the central 
galaxy $M_{vir}/M_*$ is of particular interest.
The correlation 
$M_{vir} = 1.12 \times 10^{14} (L_B/10^{11}L_{B\odot})^{1.33}$
$M_{\odot}$ shown in Figure 2a (dashed line), 
when combined with the stellar mass to light ratio 
$M_*/L_B = 7.1(L_B/10^{10})^{0.29}$ 
(Trujillo et al. 2004), results in 
$M_{vir}/M_* \approx 80 (M_*/10^{12}M_{\odot})^{0.03}$. 
This ratio is larger than the dynamical masses determined 
from the Sloan Digital Sky Survey: 
$M_{vir}/M_* \sim 15-20$ (based on $M_{vir}/L_B$ 
from Prada et al. 2003) 
and $M_{vir}/M_* \sim 7-30$ (Padmanabhan et al. 2004)
for E galaxies comparable to those in Table 1.
This suggests that 
optically similar elliptical galaxies with larger $L_x$ may 
have somewhat more massive dark halos.
%
%The dashed line in Figure 2a is an extrapolation of 
%the $L_B - M_{vir}$ relationship from 
%Lin \& Mohr (2004) determined from 2MASS data for clusters 
%with $10^{14} \lta M_{vir} \lta 10^{15}$ $M_{\odot}$, 
%adopting $L_B = 0.15 L_K$.
%The scatter around the correlation line in Figure 2a  
%is consistent with the similar cosmic scatter in 
%the Lin-Mohr correlation.
The mean gas temperature for these groups plotted 
in Figure 2b also correlates with optical luminosity of the 
group-centered galaxy as indicated by the dotted 
line with slope $T \propto L_B^{0.60}$.
Since $L_x \propto L_B^{2}$ from the Table 1 data in Figure 1,
both $M_{vir}$ and $T$ also correlate with $L_x$, 
$M_{vir} \propto L_x^{0.7}$ and $L_x  \propto T^{3.3}$. 

%The isolated point in the lower left in Figure 2a 
%for NGC 6482 may be spurious.
%$M_{vir}$ for this galaxy was determined from 
%{\it Chandra} observations that extended 
%only to 30 kpc so the small virial mass 
%and abnormally large NFW concentration ($c = 60$) probably refer 
%to the (de Vaucouleurs) stellar mass, not to that of the dark 
%matter. 
%Increasing $M_{vir}$ for NGC 6482 by $\sim 10$ 
%would move this point closer to the other galaxies 
%in Figure 2a.
%This conjecture is supported by Figure 2b where the 
%low gas temperature of NGC 6482 is a more natural extension 
%of the data for the other galaxies.

The X-ray luminosity of baryonically closed groups 
($M_{gas} \propto M_{vir}$)
scales approximately as $L_x \propto \langle n_e \rangle M_{gas}
\propto M_{vir}^2/r_{vir}^3 \propto M_{vir}$ 
[For the mean gas temperatures of the groups in Table 1, 
$1 \lta kT \lta 3$ keV, the bolometric X-ray emissivity  
is insensitive to temperature (Sutherland \& Dopita 1993)] . 
However, the $L_x - M_{vir}$ relation also depends on the 
NFW concentration 
$c = 450 (M_{vir}/M_{\odot})^{-0.128}$
(Bullock et al. 2001). 
We estimate $L_x(c,M_{vir})$ 
for closed groups ($f_g = f_b - f_* \approx 0.14$) by filling 
NFW potentials with isothermal gas with 
$T = 2.22 \times 10^7 (M_{vir}/10^{14}M_{\odot})^{0.54}$ K 
taken from Shimizu et al. (2003).
The bolometric X-ray luminosity within $r_{vir}$
is $L_x = 4.9 \times 10^{43} (M_{vir}/10^{14} M_{\odot})^{1.3}$
erg s$^{-1}$ for $10^{13.5} \lta M_{vir} \lta 10^{14.5}M_{\odot}$, 
where we assume $M_{vir} \propto L_B^{1.33}$ 
from Figure 2a (dashed line).
This locus of maximum X-ray luminosity,
$L_x = 4.3 \times 10^{43} (L_B/10^{11}~L_{B\odot})^{1.75}$ 
erg s$^{-1}$, 
is shown with a dotted line in Figure 1.
If the gas temperature 
$T \propto T_{vir} \propto M_{vir}^{2/3}$ then 
$L_x \propto M_{vir}^{1.3}$ suggests that 
$L_x \propto T^{2}$ similar to Kaiser (1986).

\section{Conclusions}

We propose that galaxy groups lying near the 
upper envelope of the $L_x - L_B$ distribution in 
Figure 1 are nearly or completely baryonically closed 
boxes similar to more massive clusters. 
This conclusion is supported by the baryon fraction  
estimates listed in Table 1 and the proximity 
of the observations in Figure 1 to 
the approximate dotted line locus 
for the maximum $L_x$ expected from 
baryonically closed groups observed to $r_{vir}$.
The projected X-ray luminosity of our 
mass model for NGC 5044 
beyond $\sim 0.1 r_{vir}$ varies as 
$L_x(r_{proj}) \propto r_{proj}^{0.28}$  
so we expect that $L_x$ for galaxies with filled symbol points 
in Figure 1 will creep upward toward the dotted line 
when observed with larger apertures 
and more sensitive detectors. 
Nevertheless, 
we do not expect to find galaxy/groups in the future that
lie significantly above the filled circles and squares
in Figure 1.

Baryonically closed groups provide interesting constraints 
on the amount of non-gravitational heating 
acquired by the intragroup gas. 
To retain most or all of the gas in these groups, 
the gas heating by the central black hole 
(AGN) must be $\eta_h M_{bh} c^2 \lta 0.016 M_{bh} c^2$ 
or $\lta 1$ keV per particle, consistent with 
typical values $\sim 0.7 - 1$ keV per particle required 
to account for deviations from self-similarity in 
the $L_x - T$ plot for clusters 
(e.g. Tornatore et al. 2003; Voit 2005).
We also find that the combined energy 
of all past supernovae 
is insufficient to remove significant amounts of 
intragroup gas unless the IMF is flatter than Salpeter
(Brighenti \& Mathews 1999, 2001).

Another necessary attribute of baryonically closed groups is 
that they are spatially isolated, 
i.e. they have not lost mass by ram-stripping 
during mergers with comparable or larger systems.
Spatially isolated E galaxies and groups are of particular interest
because of the strong limits they impose 
on non-gravitational heating. 
It is therefore remarkable that some isolated E galaxies 
have much lower $L_x$ than ellipticals in baryonically 
closed groups.
For example, in Figure 1 we mark with + symbols 
two isolated E galaxies found recently 
by Reda et al. (2004) that are near the bottom of the 
distribution.
This large variation of $L_x$ for isolated Es of 
similar $L_B$ in Figure 1 may 
result from normal cosmic variance.
It would be interesting to determine 
if these and other isolated galaxies have 
unusually undermassive dark halos (allowing winds), 
if they contain more energetically active (and massive) 
black holes or if their star formation efficiencies 
are unusually large, 
since such variations could help explain why these 
isolated galaxy/groups are not baryonically closed. 
By this means it will eventually be possible to determine if 
the non-gravitational heating arises primarily 
from the central black hole. 
It would also be worthwhile to assemble 
$M_{vir}$, $\langle T \rangle$ and optical luminosities 
for all isolated E galaxies and groups 
throughout the $L_x - L_B$ plane.

Finally, we have shown that baryonically closed groups 
can inform us about the 
important relationship between the optical 
luminosity and mass of the group-centered galaxy and the mass 
of the surrounding (group) dark halo. 
The preliminary data currently available suggest that 
the dark halos are about $\sim 80$ times more massive than 
the central (non-cD) elliptical galaxy.
These conclusions can be explored further in the 
$L_x - L_B$ plane by considering 
all elliptical galaxies for which 
X-ray observations provide accurate $M_{vir}$.

\vskip.1in

Studies of the evolution of hot gas in elliptical galaxies
at UC Santa Cruz are supported by
NASA grants NAG 5-8409 \& ATP02-0122-0079 and NSF grants
AST-9802994 \& AST-0098351 for which we are very grateful.

%\end{document}

\clearpage

%\makeatletter
%\def\jnl@aj{AJ}
%\ifx\revtex@jnl\jnl@aj\let\tablebreak=\nl\fi
%\makeatother

% use the deluxetable example from /AASTeX/table.tex:
\begin{deluxetable}{lccclcl}
%\tablefontsize{\footnotesize}
\tablewidth{0pc}
\tablecaption{LUMINOUS GALAXY GROUPS\tablenotemark{a}}
\tablehead{
\colhead{galaxy} &
\colhead{$\log (L_B/L_{B\odot})$} &
\colhead{$\log L_{x,bol}$} &
\colhead{$kT$} &
\colhead{$\log M_{vir}$} &
\colhead{$f_b$} &
\colhead{Ref.\tablenotemark{b}} \\
\colhead{group} &
\colhead{} &
\colhead{(erg~s$^{-1}$)} &
\colhead{(keV)} &
\colhead{($M_{\odot}$)} &
\colhead{} &
\colhead{}
}
\startdata
NGC5044    &10.76        &42.80 &1.2      &13.60  &    $\sim 0.13$ &1\cr
RXJ1159    &11.09        &43.05 &2.2      &14.15     & $\sim 0.04$ &2\cr
WJ943.7    &10.94        &43.07 &1.7      &13.74     &
$\sim 0.15$\tablenotemark{c}  &3\cr
RXJ0419    &10.64        &42.75 &1.4      &$\sim$13.6     &
$\sim 0.03$\tablenotemark{c}  &4\cr
NGC6482    &10.47        &42.04 &0.7      &12.60     &
$\sim 0.18$\tablenotemark{c}  &5\cr
ESO3060170 &11.30        &43.82 &2.7      &14.25 &NA  &6\cr
RXJ1340    &11.02       & 43.11 &2.3      &14.33\tablenotemark{d}
     &NA  &2,10\cr
RXJ2114    &10.96       & 43.01 &2.1      &14.27\tablenotemark{d}
     &NA  &2\cr
RXJ2247    &11.13       & 43.32 &2.8      &14.45\tablenotemark{d}
     &NA  &2\cr
NGC1132    &10.73       & 42.71 &1.0      &13.52     &NA  &7,8\cr
RXJ1416    &11.24       & 44.05 &1.5      &NA       &NA &9\cr
RXJ1119    &10.36       & 41.94 &NA      &NA        &NA &9\cr
RXJ1256    &11.16       & 43.49 &NA      &NA        &NA &9\cr
RXJ1331    &10.68       & 42.48 &NA      &NA        &NA &9\cr
RXJ1552    &11.12       & 43.51 &NA      &NA        &NA &9\cr
RXJ0116    &10.76       & 42.94 &NA      &NA        &NA &9\cr
\enddata
\tablenotetext{a}{All data scaled to $H_0$ = 70 km s$^{-1}$ Mpc$^{-1}$}
\tablenotetext{b}{References:
1 Buote et al. (2004);
2 Vikhlinin, A. et al. (1999);
3 Rasmussen, J. \& Ponman, T. J. (2004);
4 Kawaharada, M. et al. (2003);
5 Khosroshahi, H. G., Jones, L. R. \& Ponman, (2004);
6 Sun, M. et al. (2004);
7 Mulchaey, J. S. \& Zabludoff (1999);
8 Gastaldello, F. et al. (2005);
9 Jones, L. R. et al. (2003);
10 Jones, L. R. et al. (2001)
}
\tablenotetext{c}{Estimated from gas mass ratio by 
scaling up by 1.11} 
\tablenotetext{d}{Reference 2 provided values of 
the total mass $M(r_x)$ 
within radius $r_x$. We used the definition 
of $r_{vir}(M_{vir})$ with $c(M_{vir})$ and $y_x = c r_x/r_{vir}$ 
to determine $M_{vir} = M(r_x) f(c)/f(y_x)$ 
from the NFW mass profile where 
$f(x) = \ln(1+x) - x/(1+x)$.}
\end{deluxetable}

%\end{document}

%****************
\clearpage
\begin{figure}%1
\centering
\includegraphics[bb=90 216 522 569,scale=0.9,angle= 270]
{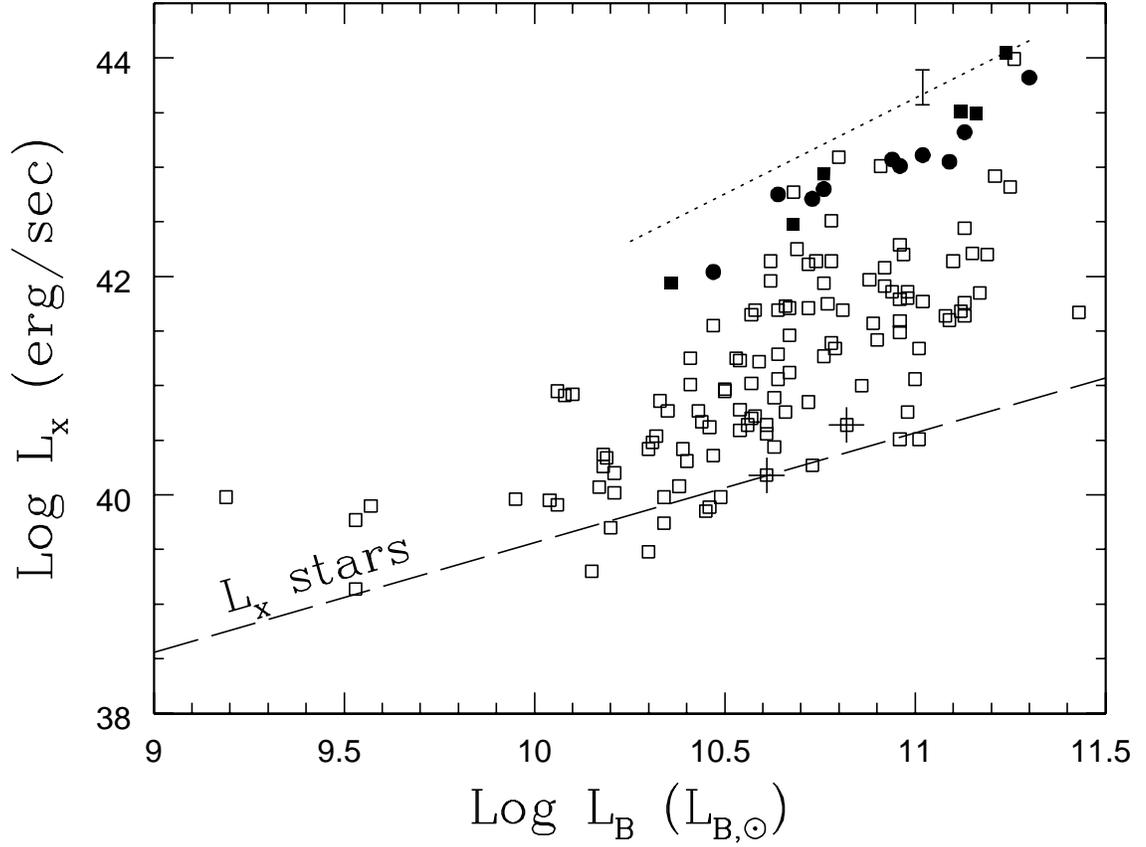}
\vskip.7in
\caption{
Plot of the bolometric X-ray luminosity and B-band 
optical luminosity for elliptical galaxies 
(RC3 type $T \le -4$) from 
O'Sullivan et al. (2001) (open squares). 
Only X-ray detected galaxies are shown.
The filled circles (squares) are X-ray luminous ellipticals 
with estimated (unknown) virial masses as listed in Table 1.
Two isolated elliptical galaxies 
(NGC 3557 and NGC 4697) are marked with + symbols.
The dashed line approximately represents the 
stellar X-ray emission from binary stars
(Kim \& Fabbiano 2004). 
The dotted line 
$L_x = 4.3 \times 10^{43} (L_B/10^{11}~L_{B\odot})^{1.75}$ 
is the locus of maximum $L_x$ 
for NFW halos maximally filled with gaseous 
baryons ($f_g = 0.14$); the error bars show 
the effect of 
a 1$\sigma$ change in concentration 
$c(M_{vir})$ expected from cosmic variation 
(Bullock et al. 2001). 
On average the filled symbols lie $\sim 0.33$ below 
the dotted line; since $L_x \propto f_g^2$ 
and the observed $L_x < L_x(r_{vir})$, they 
have gas filling factors $f_g \gtrsim 0.08$.
}
\label{fig1}
\end{figure}

\clearpage
% this was figure 3 in an earlier version of ms.
\begin{figure}%2
\includegraphics[bb=90 216 522 569,angle= 270]
{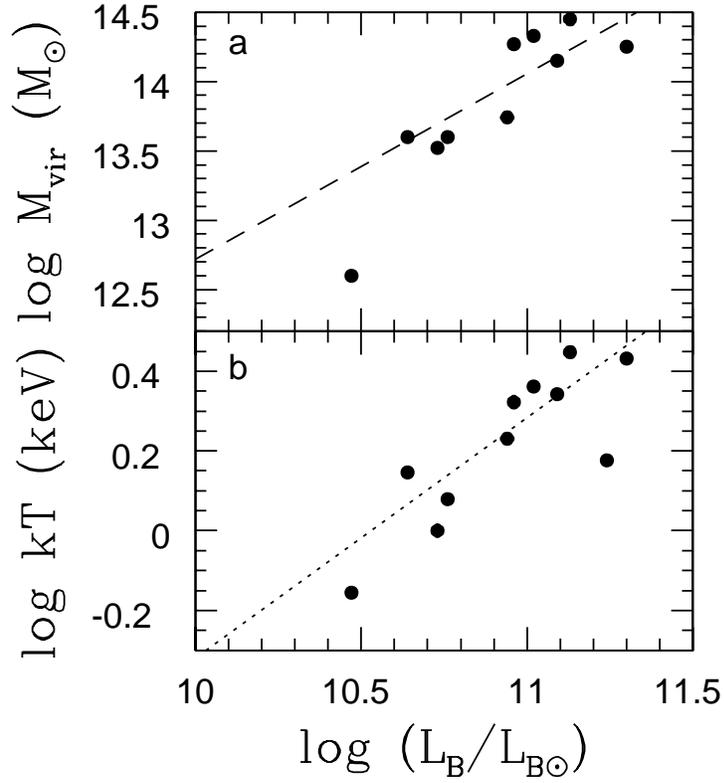}
\vskip.7in
\caption{
Plots of optical luminosity of baryonically closed 
groups against virial mass and mean gas temperature.
(a) The dashed line in the upper panel 
shows the variation $M_{vir} \propto L_B^{1.33}$ 
based on weak lensing. 
%The dotted line of slope
%$M_{vir} \propto L_B^{1.5}$ follows 
%from $L_x \propto L_B^{2}$ based on the 
%filled data points in Figure 1.
(b) The dotted line shows the correlation 
$T \propto L_B^{0.60}$.
}
\label{fig2}
\end{figure}

\end{document}